\documentclass[twocolumn, prl, a4paper, 10pt, superscriptaddress, nopacs, floatfix]{revtex4-1}
\usepackage{siunitx}
\usepackage{graphicx}
\usepackage{xcolor}
\usepackage{physics}
\usepackage{nicefrac}
\usepackage{float}
\usepackage[utf8]{inputenc}
\usepackage[colorlinks, citecolor=blue, linkcolor=black]{hyperref}
\usepackage{ulem}
\usepackage{booktabs}
\usepackage{multirow}
\usepackage{xspace}

\newcommand{\CT}{CrTe\textsubscript{2}\xspace}
\newcommand{\VB}{V$_\text{B}^-$\xspace}

\begin{document}

\title{Magnetic imaging with spin defects in hexagonal boron nitride}

\author{P. Kumar}
\author{F. Fabre}
\author{A. Durand}
\author{T. Clua-Provost}
\affiliation{Laboratoire Charles Coulomb, Université de Montpellier and CNRS, 34095 Montpellier, France}
\author{J. Li}
\author{J.~H.~Edgar}
\affiliation{Tim Taylor Department of Chemical Engineering, Kansas State University, Manhattan, Kansas 66506, USA}
\author{N.~Rougemaille}
\author{J.~Coraux}
\affiliation{Univ. Grenoble Alpes, CNRS, Grenoble INP, Institut Néel, 38000 Grenoble, France}
\author{X.~Marie}
\author{P. Renucci}
\author{C. Robert}
\affiliation{Université de Toulouse, INSA-CNRS-UPS, LPCNO, 135 Avenue Rangueil, 31077 Toulouse, France}
\author{I. Robert-Philip}
\author{B. Gil}
\author{G. Cassabois}
\author{A. Finco}
\author{V. Jacques}
\affiliation{Laboratoire Charles Coulomb, Université de Montpellier and CNRS, 34095 Montpellier, France}

\begin{abstract}
Optically-active spin defects hosted in hexagonal boron nitride (hBN) are promising candidates for the development of a two-dimensional (2D) quantum sensing unit. Here, we demonstrate quantitative magnetic imaging with hBN flakes doped with negatively-charged boron-vacancy (\VB) centers through neutron irradiation. As a proof-of-concept, we image the magnetic field produced by \CT, a van der Waals ferromagnet with a Curie temperature slightly above \SI{300}{\K}. Compared to other quantum sensors embedded in 3D materials, the advantages of the hBN-based magnetic sensor described in this work are its ease of use, high flexibility and, more importantly, its ability to be placed in close proximity to a target sample. Such a sensing unit will likely find numerous applications in 2D materials research by offering a simple way to probe the physics of van der Waals heterostructures.
\end{abstract} 
\date{\today}

\maketitle

Quantum sensing technologies based on solid-state spin defects have already shown a huge potential to cover the growing need for high-precision sensors~\cite{RevModPhys.89.035002}, both for fundamental research and for industrial applications. The most advanced quantum sensing platforms to date rely on optically-active spin defects embedded in three-dimensional (3D) materials~\cite{Awschalom2018}. A prime example is the nitrogen-vacancy (NV) center in diamond~\cite{DohertyReview,rondin_magnetometry_2014,Levine2019}, which has already found a wide range of applications in condensed matter physics~\cite{casola_probing_2018}, life sciences~\cite{Romana2014} and geophysics~\cite{Glenn2017}. Despite such success, NV-based quantum sensing technologies still face several limitations that mainly result from the 3D structure of the diamond host matrix. They include (i) a limited proximity between the quantum sensor and the target sample, which hampers its sensitivity, and (ii) the inability to engineer ultrathin and flexible diamond layers, which precludes an easy transfer of the quantum sensing unit onto the samples to be probed as well as its integration into complex multifunctional devices. An emerging strategy to circumvent these limitations consists in using spin defects embedded in a van der Waals crystal that could be exfoliated down to the monolayer limit~\cite{GottschollNatCom2021,tetienne2021}. Such a 2D quantum sensing foil would offer atomic-scale proximity to the probed sample together with an increased versatility and flexibility for device integration.

\begin{figure}[t]
  \centering
  \includegraphics[width = 8.6cm]{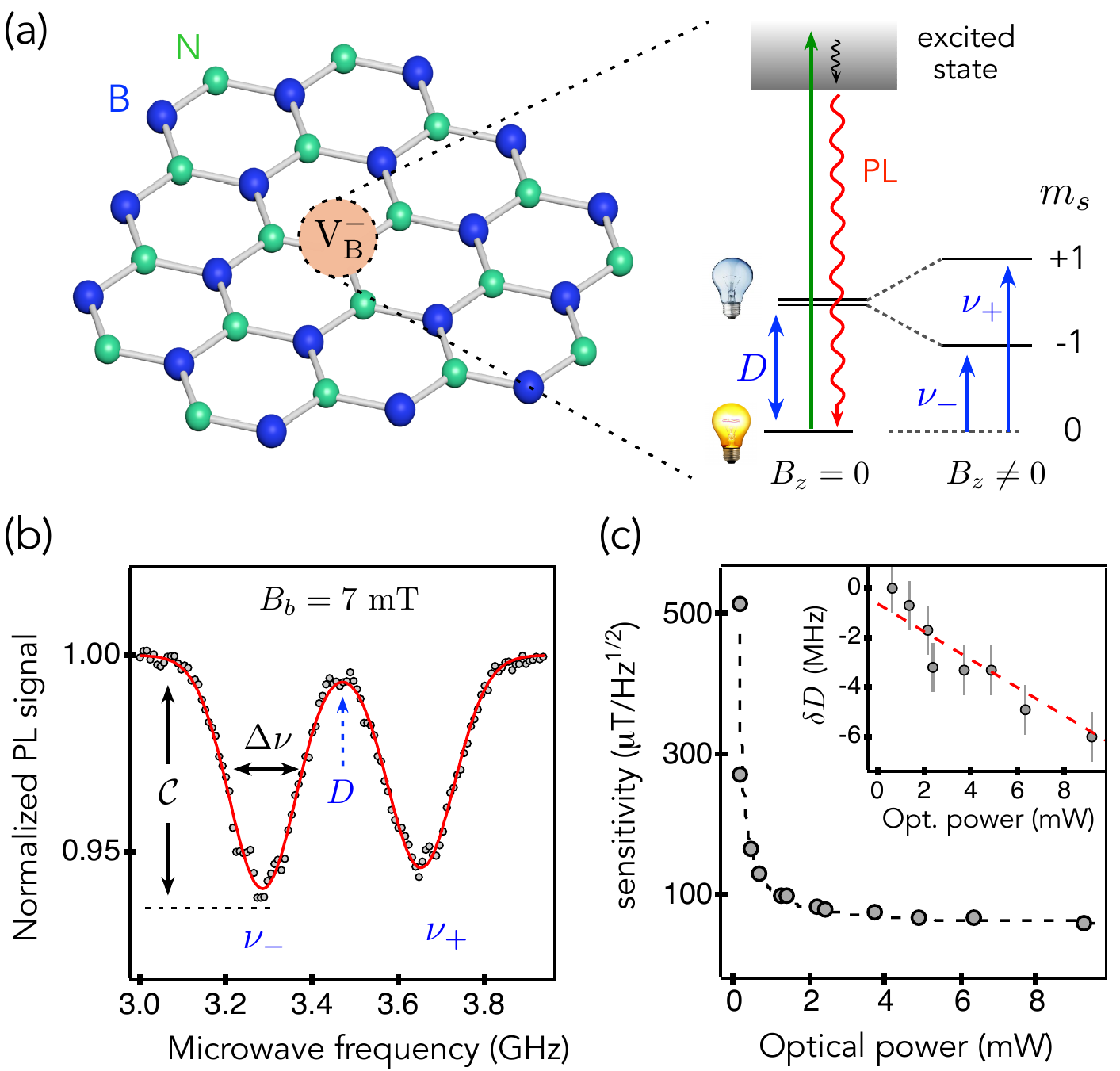}
  \caption{(a) Atomic structure of the \VB center in hBN with a simplified diagram of its energy levels showing the ESR transitions $\nu_{\pm}$ in the spin triplet ground state. (b) Optically-detected ESR spectrum recorded on a 85-nm thick hBN flake with an out-of-plane bias field $B_b=\SI{7}{\milli\tesla}$ and an optical power of \SI{2.1}{\milli\watt}. The solid line is a fit to the data with Lorentzian functions. (c) Magnetic field sensitivity $\eta_B$ as a function of the optical pumping power. The black dashed line is a guide to the eye. Inset: Evolution of $\delta D$ with the optical power. The red dashed line is a fit to the data with a linear function.}
  \label{fig1}
\end{figure}

Hexagonal boron nitride (hBN) is currently the most promising van der Waals crystal for the design of quantum sensing foils~\cite{GottschollNatCom2021,tetienne2021}. This insulating material, which can be easily exfoliated down to few atomic layers while maintaining chemical stability, is extensively used for encapsulation of van der Waals heterostructures~\cite{Geim2013}. Furthermore, hBN hosts a broad diversity of optically-active point defects owing to its wide bandgap~\cite{Tran_NatNano2016,Zobelli2016,PhysRevB.94.121405,Sajid_2020}. For some of these defects, the electron spin resonance (ESR) can be detected optically, offering a key resource for quantum sensing applications~\cite{GottschollNatCom2021}. While several spin-active defects with unknown microscopic structures have been recently isolated at the single level in hBN~\cite{Chejanovsky2021,Stern2022,Guo2021}, most studies to date have been focused on ensembles of negatively-charged boron vacancy (\VB) centers~\cite{Gottscholl2020} [Fig.~\ref{fig1}(a)]. Under green laser illumination, this defect produces a broadband photoluminescence (PL) signal in the near infrared. In addition, the \VB center features magneto-optical properties very similar to those of the NV defect in diamond, with a spin triplet ground state whose ESR frequencies can be measured via optically-detected magnetic resonance methods~\cite{Gottscholl2020,Gottscholl2021}. Despite a low quantum yield that has so far prevented its optical detection at the single scale, ensembles of \VB centers have recently found first applications as quantum sensors in van der Waals heterostructures~\cite{Tetienne2021bis,Du2021}.

In this work, we analyze the performances of hBN flakes doped with \VB centers by neutron irradiation for quantitative magnetic field imaging. As a proof-of-concept, we image the magnetic field produced by \CT, a van der Waals ferromagnet with in-plane magnetic anisotropy, which exhibits a Curie temperature ($T_{\rm c}$) slightly above \SI{300}{\K}~\cite{Freitascrte22015,Sun2020,Coraux2020,PhysRevMaterials.5.034008}. Magnetic flakes with a thickness of a few tens of nanometers were obtained by mechanical exfoliation of a bulk $1T$-\CT crystal and then transferred on a SiO\textsubscript{2}/Si substrate. For magnetic imaging, we rely on a monoisotopic h\textsuperscript{10}BN crystal grown from a Ni-Cr flux~\cite{Liu2018}, which was irradiated with thermal neutrons with a dose of about \SI{2.6e16}{n\per\centi\meter\squared}. The interest of isotopic purification with \textsuperscript{10}B lies in its very large neutron capture cross section, which ensures an efficiently creation of \VB centers via neutron transmutation doping~\cite{Li2021,haykal2021decoherence}. hBN flakes mechanically exfoliated from this neutron-irradiated crystal were deposited above the \CT flakes. Besides providing magnetic imaging capabilities, the \VB-doped hBN capping layer also protects \CT from degradation of its magnetic properties.  For each sample, the thickness of the different layers was inferred by atomic force microscopy (AFM).

All experiments described below are carried out under ambient conditions with a scanning confocal microscope employing a green laser excitation, a high-numerical aperture microscope objective (NA$=0.7$) and a photon counting detection module. At each point of the sample, optical illumination combined with microwave excitation enables the measurement of the ESR frequencies of the \VB center by recording its spin-dependent PL intensity~\cite{Gottscholl2020}. In this work, the microwave excitation is delivered via an external loop antenna placed close to the sample. 

\begin{figure*}[t]
  \centering
  \includegraphics[scale=1.05]{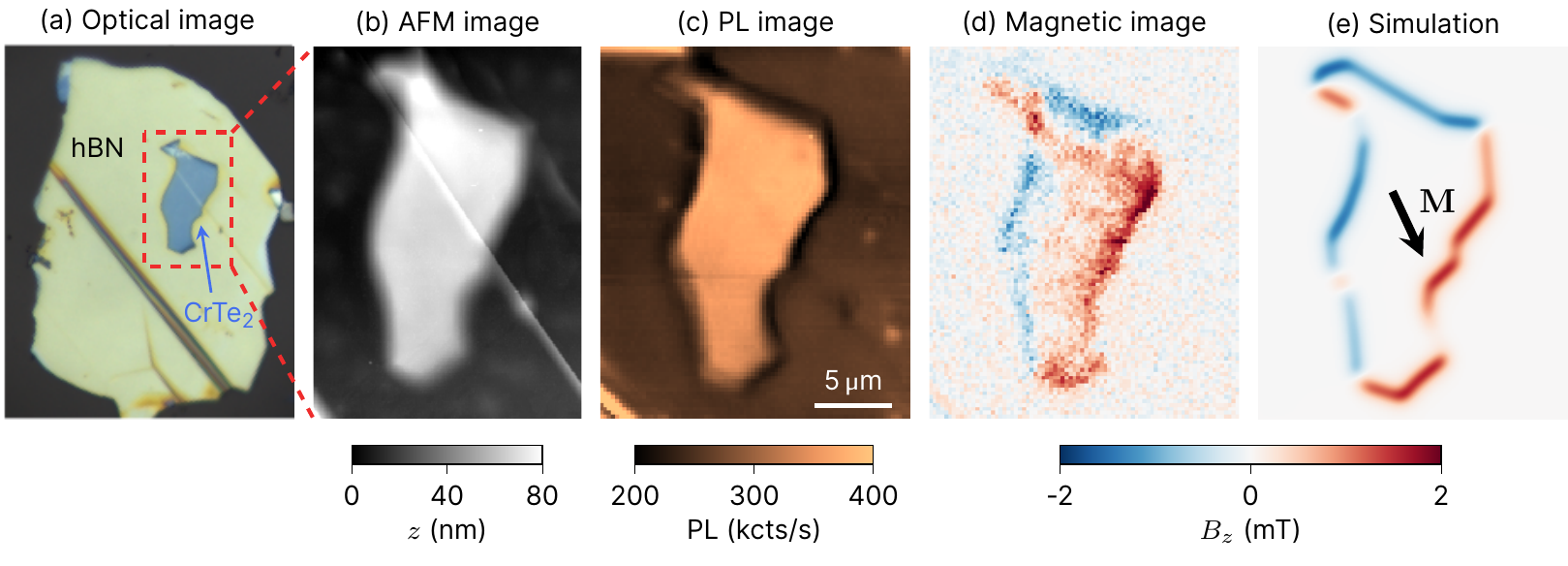}
  \caption{(a) Optical image of the \CT(\SI{64}{\nano\meter}) / hBN (\SI{85}{\nano\meter}) heterostructure. (b) AFM image indicating that the magnetic layer has a constant thickness. (c) PL image recorded across the same area as in (b).  (d) Corresponding map of the magnetic field component $B_z$ obtained by measuring the Zeeman shift of the lower ESR frequency of the \VB centers. Experiments are carried out with a bias magnetic field $B_b=\SI{7}{\milli\tesla}$ and an optical power of \SI{0.6}{\milli\watt}. The acquisition time per pixel is \SI{4.2}{\second}. (e) Simulated map of the magnetic field component $B_z$ for a uniform in-plane magnetization $\mathbf{M}$ (black arrow) with an azimuthal angle $\phi_M=\ang{297}$ and a norm $M=\SI{60}{\kilo\ampere/\meter}$.}
  \label{fig2}
\end{figure*}

We start by studying a heterostructure consisting of a 64-nm-thick \CT flake covered with a 85-nm-thick hBN sensing layer [see Fig.~\ref{fig2}(a)]. Before discussing magnetic imaging results, we first qualify the magnetic field sensitivity of the \VB-doped hBN layer as a function of the optical excitation power. To this end, optically-detected ESR spectra were recorded far from the \CT flake while applying a bias magnetic field $B_b=\SI{7}{\milli\tesla}$ along the $c$-axis of hBN ({\it i.e.} perpendicular to the layers). A typical spectrum is shown in Fig.~\ref{fig1}(b). The ESR frequencies are given by $\nu_{\pm}= D\pm \gamma_eB_b$, where $\gamma_e=\SI{28}{\GHz/\tesla}$ is the electron gyromagnetic ratio and $D\sim \SI{3.47}{\GHz}$ denotes the zero-field splitting parameter of the \VB spin triplet ground state [Fig.~\ref{fig1}(a)]. From such a spectrum, the magnetic field sensitivity $\eta_B$ can be inferred as~\cite{PhysRevB.84.195204} 
\begin{equation}
\eta_B\approx 0.7 \times \frac{1}{\gamma_e}\times\frac{\Delta\nu}{\mathcal C\sqrt{\mathcal R}} \ ,
\label{sensib}
\end{equation}
where $\mathcal{R}$ is the rate of detected photons, $\mathcal{C}$ the contrast of the magnetic resonance and $\Delta\nu$ its linewidth [Fig.~\ref{fig1}(b)]. From the parameters $\{\mathcal{R}, \mathcal{C}, \Delta\nu\}$  measured at different optical powers, we extract the power-dependent magnetic field sensitivity. As shown in Fig.~\ref{fig1}(c), $\eta_B$ improves with increasing laser power and reaches an optimal value of $\sim \SI{60}{\micro\tesla\per\sqrt{\Hz}}$. Note that in the considered power range, the PL signal does not reach saturation and the ESR linewidth $\Delta\nu$ remains almost constant (see SI). The magnetic sensitivity is therefore mainly limited by the ESR contrast $\mathcal{C}$, which could be improved by optimizing the orientation of the microwave magnetic field, {\it e.g.} by depositing the hBN flakes on top of a coplanar wave\-guide~\cite{NanoLett2021}. 

Interestingly, the gain in sensitivity with optical power is accompanied by a slight reduction of the zero-field splitting parameter $D=(\nu_++\nu_-)/2$, whose relative shift $\delta D$ is plotted in the inset of Fig.~\ref{fig1}(c). This shift results from a temperature variation $\delta T$ of the hBN layer~\cite{GottschollNatCom2021,ACSPhot_Guo2021}, which can be phenomenologically described near room temperature by the relation $\delta D= \varepsilon \ \delta T$ where $\varepsilon\sim \SI{-0.7}{\MHz/\K}$~\cite{GottschollNatCom2021}. Our measurements thus indicate that optical illumination leads to heating of the hBN sensing layer. Fitting the data with a linear function leads to an optically-induced heating efficiency around \SI{1}{\K/\milli\watt}. To mitigate this effect, which is detrimental for studying the magnetic order in materials featuring a $T_{\rm c}$ near room temperature like \CT, magnetic imaging experiments were carried out at low optical power ($\sim\SI{0.6}{\milli\watt}$), leading to a slightly degraded magnetic field sensitivity $\eta_B\sim \SI{130}{\micro\tesla\per\sqrt{\Hz}}$ [Fig.~\ref{fig1}(c)].

\indent A PL raster scan of the \CT(\SI{64}{\nano\meter}) / hBN (\SI{85}{\nano\meter}) heterostructure is shown in Fig.~\ref{fig2}(c). The increased PL signal above the \CT flake is related to its metallic character leading to reflection effects~\cite{NanoLett2021}. Magnetic field imaging is performed by recording an ESR spectrum at each point of the scan, from which the Zeeman shift of the lower ESR frequency $\nu_-$ is extracted. Here, we only track one magnetic resonance of the \VB center to reduce data acquisition time. After subtracting the offset linked to the bias magnetic field $B_b$, the Zeeman shift is simply given by $\Delta_z=\gamma_e B_z$, where $B_z$ is the magnetic field projection along the \VB quantization axis, which corresponds to the $c$-axis of hBN. A map of the magnetic field component $B_z$ produced by the \CT flake is shown in Fig.~\ref{fig2}(d). The magnetic field is mainly generated at the edges of the flake as expected for a uniformly magnetized flake with homogeneous thickness~\cite{PhysRevMaterials.5.034008}. Furthermore, the recorded magnetic field distribution directly confirms that the \CT flake features an in-plane magnetization. Indeed, considering a uniform out-of-plane magnetization, the magnetic field component $B_z$ would be identical at all edges of the flake (see SI). 

To perform a quantitative analysis of the recorded magnetic map, we simulate the distribution of the magnetic field component $B_z$ produced by a \CT flake with constant thickness and uniform in-plane magnetization $\mathbf{M}$. This vector is characterized by its azimuthal angle $\phi_M$ in the ($x, y$) sample plane and norm $M$. The geometry of the flake used for the simulation is extracted from a topography image recorded by AFM [Fig.~\ref{fig2}(b)]. By comparing the overall structure of the experimental magnetic image with simulations obtained for various values of the angle $\phi_M$, we first identify the magnetization orientation $\phi_M \sim  \ang{297}$ [Fig.~\ref{fig2}(e)]. Considering solely shape anisotropy, the in-plane magnetization should be stabilized along the long axis of the \CT flake. Our result indicates a deviation from this simple case, which suggests that \CT exhibits a non-negligible magnetocrystalline anisotropy, in agreement with recent works~\cite{Coraux2020,PhysRevMaterials.5.034008}. 

Having identified the orientation of the magnetization vector, we then estimate its norm $M$ by analyzing the stray field amplitude recorded at the edges of the \CT flake. Besides being linked to the magnetization norm, the Zeeman shift of the ESR frequency measured at each pixel of the scan also results from both (i) a lateral-averaging due to the diffraction-limited spatial resolution of the magnetic microscope and (ii) an averaging over the vertical ($z$) distribution of optically-active \VB centers in the hBN sensing layer. Indeed, neutron irradiation creates \VB centers throughout the hBN volume, in contrast to ion implantation techniques which lead to the creation of defects at a depth linked to the ion implantation energy~\cite{IgorImplant2020,ACSomega2022}. Taking into account these two averaging processes (see SI), a fair agreement is obtained between the simulated and experimental magnetic maps for $M \sim \SI{60}{\kilo\ampere/\meter}$ [Fig.~\ref{fig2}(e)]. This value is two times larger than the one recently measured for micron-sized \CT flakes without hBN encapsulation~\cite{PhysRevMaterials.5.034008}. This discrepancy is attributed to a partial degradation of non-encapsulated \CT flakes resulting in a reduced effective magnetization. As already indicated above, the magnetic image shown in Fig.~\ref{fig2}(d) was recorded at a low optical excitation power to mitigate heating effects. The same experiments performed at larger optical powers show a weaker stray magnetic field at the edges of the flake (see~SI). Since \CT has a $T_{\rm c}$ close to room temperature, any slight heating of the sample significantly reduces the sample magnetization. 

\begin{figure}[t]
  \centering
  \includegraphics[scale=1]{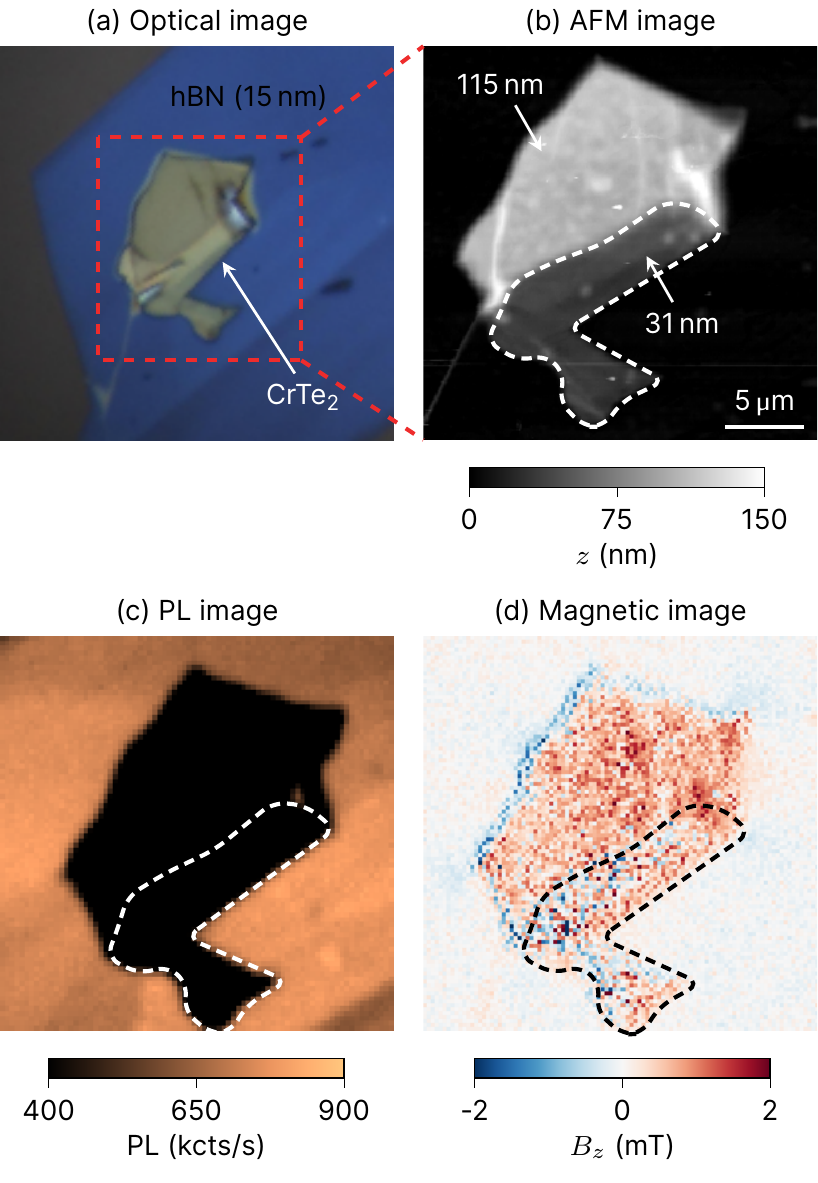}
  \caption{(a) Optical image of an heterostructure consisting of a 15-nm thick hBN layer deposited on top of a \CT flake.  (b) AFM image of the \CT flake showing large thickness variations. (c) PL image recorded around the \CT flake. (d) Corresponding map of the magnetic field component $B_z$ obtained by measuring the Zeeman shift of the lower ESR frequency of the \VB centers. Experiments are carried out with a bias magnetic field $B_b=\SI{7}{\milli\tesla}$ and an optical power of \SI{2.4}{\milli\watt}. The acquisition time per pixel is \SI{15.3}{\second}.}
  \label{fig3}
\end{figure}

Next we study a heterostructure involving a thinner (\SI{15}{\nano\meter}) hBN sensing layer in order to rely on an ensemble of \VB centers localized closer to the magnetic sample, thus reducing vertical averaging effects [Fig.~\ref{fig3}(a)]. While the PL signal increases for a thick hBN layer deposited on \CT owing to reflection effects, the PL scan of this second heterostructure now reveals a strong PL quenching of the \VB centers located above the magnetic layer [Fig.~\ref{fig3}(c)]. This effect also results from the metallic character of \CT. Indeed, when the \VB centers are placed in close proximity to the \CT flake, energy transfer to the metal quenches their PL signal by opening additional non-radiative decay channels~\cite{PhysRevLett.95.063003,TislerNanolett2013}. Even though such a PL quenching impairs the magnetic sensitivity of the hBN layer, magnetic imaging can still be performed by increasing the optical pumping power [Fig.~\ref{fig3}(d)]. Despite an improved proximity of the \VB centers, the overall amplitude of the stray magnetic field is not stronger than the one previously measured with a thicker hBN layer. This is due to laser-induced heating of the magnetic flake, which reduces its magnetization, as discussed above. For this sample, a quantitative analysis of the stray field distribution is a difficult task because of large variations of the \CT thickness. Indeed, a topography image recorded by AFM shows that the magnetic flake is divided into two main parts whose average thickness varies from $\sim$ \SI{115}{\nano\meter} to $\sim$ \SI{31}{\nano\meter}, with local fluctuations exceeding 20\% due to wrinkles [Fig.~\ref{fig3}(b)]. Since magnetic stray fields are produced at each thickness step of the magnetic layer with a possible reorientation of the in-plane magnetization~\cite{PhysRevMaterials.5.034008}, precise magnetic simulations can hardly be performed. Considering a simplified geometry of the \CT flake with a constant thickness (\SI{115}{\nano\meter}), the magnetic field produced at the top-left edge is reproduced for a magnetization $M \sim 40$~kA/m. Although qualitative, this analysis confirms the reduction of the magnetization induced by laser-induced heating of the magnetic layer.

In summary, we have shown that \VB spin defects hosted in hBN layers can be used for quantitative magnetic imaging with a sensitivity around $\SI{100}{\micro\tesla\per\sqrt{\Hz}}$ and a spatial resolution limited by diffraction at the micron scale. Although much better performances can be obtained with other quantum sensors such as the NV defect in diamond, the key advantages of the hBN-based magnetic sensor described in this work are its ease of use, high flexibility and, more importantly, its ability to be placed in close proximity to a target sample. Such a sensing unit will likely find numerous applications in 2D materials research by offering a simple way to probe {\it in situ} the physics of van der Waals heterostructrures, with optimal performance obtained for the study of non-metallic 2D materials, for which PL quenching effects can be avoided. An improvement in magnetic sensitivity by at least one order of magnitude could be achieved by optimizing the microwave excitation used to perform ESR spectroscopy~\cite{NanoLett2021} while spatial resolution below the diffraction limit might be reached by relying on super-resolution optical imaging methods~\cite{Kianinia2018,Radenovic2019,Luxmoore2021}. To release the full potential of hBN-based quantum sensing foils, a remaining challenge, however, is to demonstrate that \VB centers can be stabilized in an atomically-thin hBN layer to achieve an ultimate atomic-scale proximity with the probed sample.

\noindent {\it Acknowledgements} - This work was supported by the French Agence Nationale de la Recherche under the program ESR/EquipEx+ (grant number ANR-21-ESRE-0025), the Institute for Quantum Technologies in Occitanie through the project BONIQs, an Office of Naval Research award number N000142012474 and by the U.S. Department of Energy, Office of Nuclear Energy under DOE Idaho Operations Office Contract DE-AC07-051D14517 as part of a Nuclear Science User Facilities experiment. We acknowledge Abdellali Hadj-Azzem for providing the bulk $1T$-CrTe$_2$ samples and the support of The Ohio State University Nuclear Reactor Laboratory and the assistance of Susan M. White, Lei Raymond Cao, Andrew Kauffman, and Kevin Herminghuysen for the irradiation services provided. 

\noindent {\it Data Availability Statement} -
The data that support the findings of this study are openly available in Zenodo at https://zenodo.org/record/6802738 with the identifier 10.5281/zenodo.6802738.


\end{document}